
\input harvmac 
\def\tr{{\rm tr\,}}
\def\ud{\half}
\def\ee#1{{\rm e}^{^{\textstyle#1}}}
\def\d{{\rm d}}
\def\e{{\rm e}}

\def\pp{r}
\def\qq{s}
\def\rr{\pp}
\def\Om{\Omega}
%
\Title{\vbox{\baselineskip12pt\hbox{SPhT/92-042}}}
\centerline{\titlefont The $O(n)$ model on a random surface: critical points
and large order behaviour}
\vskip1.3truecm
\centerline{B. Eynard and J. Zinn-Justin$^1$}
{\baselineskip14pt\centerline{Service de Physique Th\'eorique$^2$ de Saclay}
\centerline{F-91191 Gif-sur-Yvette Cedex, FRANCE}}
\footnote{}{$^1$zinn@poseidon.saclay.cea.fr}
\footnote{}{$^2$Laboratoire de la Direction des Sciences de la Mati\`ere du
Commissariat \`a l'Energie Atomique}
\vskip0.3truecm
In this article we report a preliminary investigation of the large $N$ limit
of a generalized one-matrix model which represents an
$O(n)$ symmetric model on a random lattice. The model on a regular lattice is
known to be critical only for $-2\le n\le 2$. This is the situation we shall
discuss also here, using steepest descent. We first determine the critical
and multicritical points, recovering in particular results previously
obtained by Kostov. We then calculate the scaling behaviour in the critical
region when the cosmological constant is close to its critical value. Like
for the multi-matrix models, all critical points can be classified in terms
of two relatively prime integers $p,q$. In the parametrization $p=(2m+1)q \pm
l$, $m,l$ integers such that $0<l<q$, the string susceptibility exponent is
found to be $\gamma_{\rm string}=-2l/(p+q-l)$. When $l=1$ we find that all
results agree with those of the corresponding $(p,q)$ string models,
otherwise they are different.\par
We finally explain how to derive the large order behaviour of the
corresponding topological expansion in the double scaling limit.
\Date{22/04/92}
\nref\rIK{I. Kostov, {\it Mod. Phys. Lett.} A4 (1989) 217, {\it Phys. Lett.}
B266 (1991) 312\semi M. Gaudin and I.Kostov, {\it Phys. Lett.} B220 (1989)
200.}
\newsec{Introduction}
We discuss in this article an $O(n)$ symmetric matrix model in the large
$N$ limit. The model has a direct interpretation in
terms of a gas of loops \rIK, characterized by an $O(n)$ index, drawn on
Feynman  diagrams. The partition function is initially
given by an integral over $n+1$ matrices, but the integral over $n$ of them
is gaussian. The gaussian integrations generate an unusual effective
one-matrix model which, unlike the standard one-matrix model
\ref\rDavidetal{F. David, {\it Nucl. Phys.} B257[FS14] (1985) 45, 543\semi
J. Ambj{\o}rn, B. Durhuus and J. Fr\"ohlich, {\it Nucl. Phys.} B257[FS14]
(1985) 433; J. Fr\"ohlich, in: {\it Lecture Notes in Physics,} Vol. 216,
ed. L. Garrido (Springer, Berlin, 1985)\semi
V. A. Kazakov, I. K. Kostov and A. A. Migdal, {\it Phys. Lett.} 157B (1985)
295; D. Boulatov, V. A. Kazakov, I. K. Kostov and A. A. Migdal,
{\it Phys. Lett.} B174 (1986) 87; {\it Nucl. Phys.} B275[FS17] (1986) 641.},
cannot be
solved by the standard orthogonal polynomial method \ref\ronemat{M. Douglas
and S. Shenker, {\it Nucl. Phys.} B335 (1990) 635\semi
E. Br\'ezin and V. Kazakov, {\it Phys. Lett.} B236 (1990) 144\semi
D. Gross and A. Migdal, {\it Phys. Rev. Lett.} 64 (1990) 127;
{\it Nucl. Phys.} B340 (1990) 333.}.\par
The resulting model is solved instead, in the spherical limit, by the method
of steepest descent. The saddle point equations generalize similar equations
for the usual one-matrix model, but the method of solution involves new
technical considerations. We show in particular that when $n$ is of the form
$n=-2\cos(\pi p/q)$, with $p,q$ positive relatively prime integers, the trace
of the resolvent of the matrix is the solution
of an algebraic equation of degree $q$ with polynomial coefficients.\par
As expected from the analysis of two dimensional regular lattices,
non-trivial critical behaviour is only found for $-2\le n \le 2$. We exhibit
a class of critical and
multicritical points, recovering in particular results previously derived by
Kostov \refs{\rIK}. We mainly discuss the case $n=-2\cos(\pi
p/q)$, $p,q$ being two relatively prime integers, although some
considerations also apply to arbitrary values of $n$. When $p$ has the form
$p=(2m+1)q\pm 1$ we find results consistent with the $(p,q)$ models (in the
CFT classification  \ref\rBPZ{A. A. Belavin, A. M.
Polyakov and A. B. Zamolodchikov, {\it Nucl. Phys.} B241 (1984) 333.})  as
previously obtained in the multi-matrix models \ref\rGGPZ{P. Ginsparg, M.
Goulian, M. R. Plesser, and J. Zinn-Justin,
{\it Nucl. Phys.} B342 (1990) 539.}. This includes the
unitary family. However in general the scaling properties are different
(or at least the operator content).
This is not totally surprising since in general these models cannot be
directly related to multi-matrix models.
There is however one exception, the two-matrix model representing the Ising
model on a random triangular lattice. This model is identical to an $O(1)$
model with a special cubic potential. It will be investigated first in order
to show that the results obtained by the method of
orthogonal polynomials \ref\rIsing{E. Br\'ezin, M. Douglas, V. Kazakov, and
S. Shenker, {\it Phys. Lett.} B237 (1990) 43\semi
D. Gross and A. Migdal, {\it Phys. Rev. Lett.} 64 (1990) 717\semi
C. Crnkovi\'c, P. Ginsparg, and G. Moore, {\it Phys. Lett.}
B237 (1990) 196.} are recovered at leading order.\par
We then discuss multicritical points in a general potential in the $O(1)$
model. We finally study the critical points of the general $O(n)$ models.
\par
For the one-matrix model the large order behaviour of the topological
expansion has first been determined by a direct analysis of the
differential equations \nref\rGZlob{P. Ginsparg and J.
Zinn-Justin, 
{\it Phys. Lett.} B255 (1991) 189.}%
\nref\rGZaplob{P. Ginsparg and J. Zinn-Justin,
``Action principle and large order behaviour of non-perturbative gravity'',
LA-UR-90-3687/SPhT/90-140 (1990), to appear in {\it Random surfaces and
quantum gravity\/}, proceedings of 1990 Carg\`ese workshop, edited by
O. Alvarez, E. Marinari, and P. Windey.} \refs{\rGZlob,\rGZaplob} satisfied
by the scaling specific heat. It has been shown that perturbation series is
divergent and for half of the models (this includes pure gravity) non-Borel
summable. This property has later been
related to the existence of non-trivial saddle points of the initial matrix
integral in the large $N$ limit \ref\rFDii{F. David, {\it Nucl. Phys.} B348
(1991) 507.} (the equivalent of the instanton solution
of quantum mechanics and field theory). In the non-Borel summable case these
saddle points are real, a property related to the unbounded nature of the
integrand in the scaling limit or equivalently to the instability of the
corresponding statistical models by creation of surfaces of higher
topologies.\par
In the case of the multi-matrix model, the large order behaviour of
the critical and tricritical Ising models has been completely determined,
again by a direct analysis of the differential equations. For the general
$(p,q)$ model only the $2k!$ divergence of perturbation theory has been
established. The application of the steepest descent method to
the multi-matrix integrals instead is not easy, because the leading
contributions cancel since now the square of the Vandermonde determinant is
replaced by the product of two independent determinants involving the
eigenvalues of the first and last matrices of the chain. \par
We show that the models considered here, because they can be solved by
steepest descent, allow for a direct determination of the large order
behaviour. \par
Since the discussion of the critical points involves many technical
considerations, in this article we explain the methods, establish the
scaling of the free energy, explicitly calculate the trace of the resolvent
of the matrix in the scaling limit, and compare it to the resolvent of the
$(p,q)$ string models. In a next article we shall examine the
critical behaviour in more details.
\newsec{The critical Ising model}
We first consider a special example of the two-matrix model with
the partition function given by
\eqn\epartZI{Z=\int\d M_1\d M_2\, \ee{-(N/g)\tr U(M_1,M_2)},}
$M_1,M_2$ being two $N\times N$ hermitian matrices and the potential $U$
having the form
\eqn\epotI{U(M_1,M_2)=\ud(M_1^2+M_2^2)-cM_1 M_2+(M_1^3+M_2^3)/3,\quad 0<c<1
.}
This model can of course be exactly solved by the method of orthogonal
polynomials \rIsing. We expect that it exhibits a double scaling limit
representing the critical Ising model on a random surface. We solve it
however here by a
different method and only at leading order for $N$ large. A comparison with
the large $N$ limit of the exact result provides a check for the new
method of solution which we can then apply to the general $O(n)$ model.\par
We first transform integral \epartZI, setting
$A=(M_1-M_2)/2$ and $S=M_1+M_2+1+c$. Substituting into the potential \epotI\
we find
$$\tr U(M_1,M_2)=\tr [A^2S+V(S)],$$
with
\eqn\epotS{4V =S^3/3-2cS^2+(3c-1)(1+c)S+{\rm const.}.}
Eventually the matrix $A$ will be replaced by $n$ copies,
the Ising model being identified with the $O(1)$ case of the $O(n)$ model.
\par
The gaussian integral over $A$ can be performed and $Z$ is now given by an
integral over only one matrix $S$:
\eqn\eZIii{Z=\int\d S\, [\det(S\otimes 1)]^{-1/2}\e^{-(N/g)\tr V(S)}.}
The integral over unitary matrices can still be performed, and the result is
\eqn\eZIiii{Z=\int \Delta^2(\Lambda) \prod_{i=1}^N \d
\lambda_i\,\e^{-(N/g)V(\lambda_i)}\prod_{i,j}
\left(\lambda_i+\lambda_j\right)^{-1/2},}
where we denote by $\Delta(\Lambda)$ the usual Vandermonde  determinant of the
eigenvalues of $S$. In this form the model can no longer be solved by the
method of orthogonal polynomials, but conversely it is much easier, as we
shall show, to solve it, in the spherical limit, by the method of
steepest descent and then to also find non-trivial saddle points.
\subsec{The saddle point equation}
Varying one eigenvalue of $S$ we obtain the saddle point equation
\eqn\esaddi{{2\over N}\sum_{j\ne i}{1 \over \lambda_i-\lambda_j}=
{1\over N}\sum_j {1\over \lambda_i+\lambda_j}+{1\over g}V'(\lambda_i),}
with
\eqn\epotIs{4V'(\lambda)=(\lambda-2c)^2-(c-1)^2=(\lambda-3c+1)(\lambda-c-1).}
In addition to the repulsive potential between eigenvalues, already present
in the one-matrix model, we have now an
attractive and unbounded potential at $-\lambda_j$. A potential barrier
must separate the eigenvalues from their reflected images and thus a
neighbourhood of the origin must be free of eigenvalues, all eigenvalues being
for $g$ positive on the positive real axis.\par
It is convenient to rewrite equation \esaddi\ in terms of the trace
$\omega_0(z)$ of the resolvent of the matrix $S$,
\eqn\ersol{\omega_0(z)={1\over N}\tr{1\over z-S}={1\over N}\sum_i {1 \over
z-\lambda_i}.}
In the large $N$ limit the distribution of eigenvalues
$\rho(\lambda)=N^{-1}\sum_i\delta(\lambda-\lambda_i)$
becomes a continuous function and equation \esaddi\ can be rewritten:
\eqn\esaddii{\omega_0(z+i0)+\omega_0(z-i0)=-\omega_0(-z)+V'(z)/g.}
The distribution $\rho(\lambda)$ is then given by
$$\rho(\lambda)=-{1\over
2i\pi}\bigl(\omega(\lambda+i0)-\omega_0(\lambda-i0)\bigr).$$
A particular polynomial solution of the equation \esaddii\ is
\eqn\esolo{\omega_r(z)={1 \over 3g}\bigl(2V'(z)-V'(-z)\bigr).}
Setting
\eqn\eomegdef{\omega_0(z)=\omega_r(z)+{1\over 12g}\omega(z),}
we then look for a one-cut solution $\omega(z)$ of the homogeneous equation
which, as a consequence of equation \epotIs\ and the definition of
$\omega_0$, for $z$ large behaves like
\eqn\eomlarg{\omega(z)=-4\bigl(2V'(z)-V'(-z)\bigr)
+(12g)/z+O\left(z^{-2}\right)\sim -z^2 .}
\medskip
{\it A quadratic relation.}
We have now to solve the homogeneous equation
\eqn\egenRH{\omega(z+i0)+\omega(z-i0)=-\omega(-z),}
with the hypothesis that $\omega(z)$ is a real analytic function
with a unique cut on the real positive axis ($z=0$ excluded). \par
This equation has an equivalent form. Let us consider the
even function
\eqn\equad{\pp(z)={\textstyle{1\over3}}
\left[\omega^2(z)+\omega^2(-z)+\omega(z)\omega(-z)\right],}
and calculate its discontinuity across the cut on $z>0$, where $\omega(-z)$ is
regular:
\eqna\esaddiii
$$\eqalignno{\pp(z+i0)-\pp(z-i0)&={\textstyle{1\over3}}
\left[\omega(z+i0)-\omega(z-i0)\right] &\cr&\quad \times
\left[\omega(z+i0)+\omega(z-i0)+\omega(-z)\right].&\esaddiii{}\cr}$$
The r.h.s.\ vanishes and thus $\pp(z)$ is an even regular function, which
taking into account the large $z$ behaviour of $\omega(z)$  is a polynomial
of degree 4.  Moreover equation \esaddiii{}\ is equivalent to equation
\egenRH.
\medskip
{\it Direct derivation.} Note that equation \equad\ can be directly derived
from equation \esaddi\ by multiplying
\esaddi\ by $1/(z-\lambda_i)$ and summing over $i$. From the two identities
$$\eqalign{{2\over N^2}\sum_{i\ne j}{1\over z-\lambda_i}{1\over
\lambda_i-\lambda_j}&= \omega_0^2(z)+{1\over N}\omega'_0(z),\cr
{1\over N^2}\sum_{i,j}\left({1\over z-\lambda_i}-{1\over z+\lambda_j}\right)
{1\over \lambda_i+\lambda_j}&=-\omega_0(z)\omega_0(-z),\cr}$$
and neglecting a term of order $1/N$, we find for $\omega_0(z)$:
$$\omega_0^2(z)+\omega_0^2(-z)+\omega_0(z)\omega_0(-z)={1\over
g}\bigl(V'(z)\omega_0(z)+V'(-z)\omega_0(-z)\bigr)+r_1(z),$$
where $r_1$ is a constant when $V'(z)$ is a polynomial of degree 2:
$$g\, r_1(z)=\int\d \lambda\,\rho(\lambda)\left[{V'(\lambda)-V'(z)\over
z-\lambda}-{V'(\lambda)-V'(-z)\over z+\lambda}\right]=2c-\int\d
\lambda\,\rho(\lambda)\lambda\,.$$
Shifting $\omega_0(z)$ we obtain an explicit expression for $\pp(z)$.
We note in particular that a variation of $g$ translates for $\omega$
only into a variation of $r_1$. Finally the same equation can also obtained
from the loop equations (the equations of motion) in the large $N$ limit.
\subsec{General one-cut solution}
{\it A cubic algebraic equation.}
Equation \egenRH\ implies that the branch points are square root branch points
because turning twice around them we return to the initial function.
However in the second sheet we find also the cut of $\omega(-z)$. If we pass
through this new cut we arrive in general onto a new sheet and so on.
Here, due to the special coefficients, we find only three different sheets,
which implies that $\omega(z)$ satisfies an algebraic equation of third
degree.  This equation can immediately be obtained by multiplying equation
\equad\ by $\omega(z)- \omega(-z)$. We then find:
\eqn\eomegiii{\omega^3(z)-3\pp(z)\omega(z)=\omega^3(-z)-3\pp(z)\omega(-z)
=2\qq(z),}
where $\qq(z)$ is an even function which is everywhere regular because the
first expression is regular for $z<0$ and the second for $z>0$. It is
therefore a polynomial.
\medskip
{\it General solution.}
The explicit solution of equation \eomegiii\ can be written
\eqna\eCardan
$$\eqalignno{\omega(z)&=\e^{-2i\pi/3}\omega_+(z)+\e^{2i\pi/3}\omega_-(z),
&\eCardan a \cr
\omega_{\pm}(z)&= \left(\qq(z)\pm\sqrt{\Delta(z)}\right)^{1/3},& \eCardan b
\cr}$$
where $\Delta$ is the discriminant of the equation
\eqn\ediscr{\Delta(z)=\qq^2(z)-\pp^3(z).}
Because we look for solutions $\omega(z)$ which are not even in $z$, we verify
that we can write $\sqrt{\Delta}$:
$$\sqrt{\Delta}=i 3^{-3/2}\bigl(\omega(z)-\omega(-z)\bigr)\bigl(
\omega^2(z)+\omega^2(-z)+{\textstyle{5\over2}}\omega(z)\omega(-z)\bigr).$$
This expression in particular shows that $\sqrt{\Delta}$ is an odd function.
It follows that $\omega_-(z)= \omega_+(-z)$ and therefore
\eqn\eompmrec{\omega_{\pm}(z)=\pm{1\over i\sqrt{3}}\left[\e^{\mp 2i\pi/3}
\omega(z)-\e^{\pm 2i\pi/3}\omega(-z)\right].}
Our normalizations are such that $\omega_{\pm} \sim z^2$.
\par
The reciprocal formulae are also useful
\eqn\erecipq{\pp(z)=\omega_+(z)\omega_-(z),\qquad
\qq(z)=\ud\left(\omega^3_+(z) +\omega^3_-(z)\right).}
It follows:
\eqn\erecD{\sqrt{\Delta}=\ud \left(\omega^3_+(z)-\omega^3_-(z)\right).}
The expansion \eomlarg\ of $\omega(z)$ for large $z$ implies that
$\sqrt{\Delta}$ behaves like $z^5$. For the general  one-cut solution
$\sqrt{\Delta}$ must then have the form
$$\sqrt{\Delta}=12\sqrt{3}ic z\left(z^2-e^2\right)\left[
\left(z^2-a^2\right) \left(z^2-b^2\right)\right]^{1/2},$$
where we have chosen $b\ge a \ge 0$. The condition that $\Delta(0)=0$ implies
the relation
\eqn\epqze{\qq^2(0)=\pp^3(0).}
The determination of $\sqrt{\Delta}$  has been chosen such that
$$\omega_{\pm}(z-i0)=\e^{\pm 2i\pi/3}\omega_{\mp}(z+i0),\quad {\rm for}\ z\in
[a,b].$$
The relation $\omega_+(z)=\omega_-(-z)$ then automatically implies that
$\omega(z)$ has in the first sheet  a cut  only along $[a,b]$. Indeed
$$\eqalign{{\rm for}\ z>0\quad\omega(z-i0)&=\omega_+(z+i0)+\omega_-(z+i0),
\cr {\rm for}\ z<0\quad\omega(z-i0)&=\omega(z+i0).\cr}$$
\medskip
{\it Critical points.} A critical point is generated by the confluence of two
different zeros of $\Delta$. Two situations can arise:\par
(i) $a=e$, this is the
case of an ordinary critical point of the one-matrix model, the determinant
coming from the integration over $A$ playing no special role. It just
modifies the form of the potential. From the point of view of the Ising model
this is the low temperature phase where all Ising degrees of freedom are
frozen. \par
(ii) $a=0$, this is a new critical point specific to the structure of integral
\eZIiii, and the only case we shall consider from now on. The condition
$a=0$ implies that $N<\tr A^2>$, which characterizes the Ising spin
fluctuations, diverges. Indeed $N<\tr A^2>$ is proportional to
$\sum_{i,j}1/(\lambda_i+\lambda_j)$ and diverges only when some eigenvalue of
$S$ vanishes \refs{\rIK}. This argument is confirmed in the continuum limit
by a determination of the scaling properties of $N<\tr A^2>$. \par
Finally the critical Ising model in
the continuum limit is reached when both confluences occur simultaneously,
$a=e=0$. Note, however, that in this limit the eigenvalue distribution
approaches a singularity of the integrand. Therefore we can no longer be
certain that the steepest descent method is valid beyond the leading order
for $N$ large. We have made a few checks but this point requires a
more detailed investigation.
\subsec{Simple critical point}
{\it General critical solutions.} Solutions corresponding to these critical
points will be characterized by the property that they have a cut only
between the origin and $z=b>0$. The functions $\omega_{\pm}$ corresponding to
these solutions have the following general form:
\eqn\eomcrgen{\omega_{\pm}= \left(\sqrt{1-b^2/z^2}\mp ib/z\right)^{1/3}
\left(A(z)\sqrt{1-b^2/z^2}\pm ibB(z)/z\right),}
where $A$ and $B$ are even polynomials. The functions
$\pp,\qq,\Delta$ have then indeed the expected analytic properties:
\eqna\ecritic
$$\eqalignno{\pp(z)&=z^{-2} \left[A^2(z)\left(z^2-b^2\right)+b^2
B^2(z)\right],&\ecritic{a}\cr
\qq(z)&=2 z^{-4} \left[A^3\left(z^2-b^2\right)^2+3\left(A^2 B-B^2 A\right)
b^2\left(z^2-b^2\right)-b^4B^3\right],&\ecritic{b}\cr
\sqrt{\Delta}&=ibz^{-4}\sqrt{z^2-b^2}\left[b^2B^2(3A-B)+\left(z^2-b^2\right)
A^2(3B-A)\right],&\ecritic{c}\cr}$$
provided $A(z)$ and $B(z)$ vanish at $z=0$. The degrees of the polynomials
$A,B$ are directly connected to the large $z$ behaviour of $\omega_{\pm}$ and
thus to the degree of the potential $V(S)$.
\medskip
{\it The cubic potential.} In the case of the potential \epotS\
$\omega_{\pm}\sim z^2$ for $z\to\infty$. These conditions imply $A(z)=z^2$
and $B(z)\propto z^2$, and therefore the
general solution \eomcrgen\ reduces to
\eqn\epgplus{\omega_{\pm}=z^2  \left(\sqrt{1-b^2/z^2}\mp ib/z\right)^{1/3}
\left(\sqrt{1-b^2/z^2}\pm ib\sigma/z\right).}
The criticality condition $a=0$ defines a line $g=g_c(c)$ in the $g,c$ plane,
and thus the two parameters $b,\sigma$ are functions of $c$.
Note that the double zeros $\pm e$ of $\Delta$ are given by
$e^2=b^2(\sigma-1)^3/(3\sigma-1)$. They vanish for $\sigma=1$ which thus
corresponds to the Ising model critical point, a case we examine
later. We first discuss the generic case $\sigma\ne 1$.
\par
To relate the parameters $b, \sigma$ with the parameters $c,g$
which characterize the special potential \epotS\ of the Ising model
we can compare the expansions of $\omega_+(z)$ for $z$ large up to order
$1/z$. Using equations \eomlarg\ and \eCardan{a} we find at the critical
point $b(3\sigma-1)=12\sqrt{3}c$, $b^2(3\sigma-5)/9=(3c-1)(1+c)$,
and
\eqn\egcpg{12 g_c\sqrt{3}=b^3\left({19\over 54} -{\sigma\over6}\right).}
The conditions $0\le c$ and $g>0$ imply $1/3\le \sigma \le 19/9$.
The limit $\sigma\to -\infty$, which cannot be reached in the Ising model,
corresponds to a quadratic potential $V(S)$. For $c=0$, i.e.\ $\sigma=1/3$ we
know the critical value of $g$ from the solution of the one-matrix model,
$g_c=1/(12\sqrt{3})$, value which agrees with the result \egcpg.
\medskip
{\it Critical behaviour.} The critical behaviour is obtained from the small
$z$ behaviour of $\omega_{\pm}$. For $\sigma\ne 1$,
$$\omega(z)\propto z^{2/3}.$$
For $\sigma=1$ the behaviour is different and is discussed in the
section 2.5. The amplitude in front of $z^{2/3}$, which is needed below
to determine the normalization of $\omega(z)$ in the scaling limit,
depends on the determinations. We therefore choose to let $z$ approach the
origin on the negative imaginary axis: $z=-i\lambda$, $\lambda\to 0_+$, a
choice we shall keep throughout the article. Then
\eqna\eomcas
$$\eqalignno{\omega_+(-i\lambda)&\sim-2^{1/3}(1-\sigma)b^{4/3}\lambda^{2/3},
&\eomcas{a}\cr
\omega_-(-i\lambda)&\sim -2^{-1/3}(\sigma+1)b^{2/3}\lambda^{4/3}.
&\eomcas{b}\cr}$$
\subsec{The scaling limit}
We now  calculate $\omega(z)$ near the critical point,
when the cosmological constant is close to its critical value, i.e.\ for
$$|x=1-g/g_c|\ll 1\,,$$
in the scaling limit. Scaling functions are solutions of equation \egenRH\
with a cut for  $z\ge a>0$. The scaling functions $\omega_{\pm}$ have
the general form
\eqn\eomscgen{\omega_{{\rm sc,}\pm}
=\left(\sqrt{a^2-z^2}\mp iz\right)^{1/3}\left(C(z)
\sqrt{a^2-z^2}\pm iz D(z)\right),}
where $D,C$ are two even polynomials whose form is fixed by comparing the
large $z$ behaviour of $\omega_{\rm sc}$ with the small $z$
expansion of the critical functions. \par
Equations \eomcas{} yield the large $z$ behaviour of $\omega_{\rm sc}$ for
$\lambda\to+\infty$ and thus the degrees of the polynomials $C,D$. Here
$C$ and $D$ must be equal constants and thus:
\eqn\eompg{\omega_{{\rm
sc},\pm}=-2^{-1/3}b^{4/3}(1-\sigma)\left(\sqrt{a^2-z^2}\pm iz\right)^{2/3} .}
It remains to obtain the relation between $a$ and $x$. One can for this
purpose directly calculate the variation of the polynomials $\pp,\qq$ at
leading order in $x$. However
we use here a different argument which can easily be generalized to more
complicate cases.
\medskip
{\it An auxiliary function.} We consider the function
$\Omega(z)$:
\eqn\eOmdef{\Omega(z)={\partial\over\partial g}\bigl(g\omega_0(z)\bigr).}
We note that \eomegdef\ implies
$$\Omega(z)={1\over12}{\partial\omega(z)\over\partial g}.$$
It thus satisfies the homogeneous equation \egenRH. It is then convenient to
introduce the corresponding two functions $\Omega_{\pm}$:
$$\Omega_{\pm}=\pm{1\over i\sqrt{3}}\left[\e^{\mp 2i\pi/3}
\Omega(z)-\e^{\pm 2i\pi/3}\Omega(-z)\right],$$
and therefore
\eqn\eOmpm{\Omega(z)=\e^{-2i\pi/3}\Omega_+(z)+\e^{2i\pi/3}\Omega_-(z).}
Since $\Omega$ is the derivative
of a function which has a singularity of the form $(z-z_0)^{1/2}$, $z_0=\pm
a, \pm b$, it can have a
stronger singularity of $(z-z_0)^{-1/2}$ type. Moreover from the definition
of $\omega_0$ we infer the behaviour of $\Omega$ for $z$ large, $\Omega(z)\sim
1/z$. These conditions determine
$\Omega(z)$ uniquely as a function only of the location $a,b$ of the
singularities, a property we shall use later. At the critical point
$\Omega_{\pm}$ have the form \eomcrgen, where however now $A(z)$ can have
poles at $z=\pm b$ and $A(z)$ and $B(z)$ no longer necessarily vanish at
$z=0$. The large $z$ behaviour of $\Omega(z)$ implies
$$B(z)\mathop{\sim}_{z\to\infty}1/(b\sqrt{3}),\quad
A(z)=\widetilde{A}(z)/(z^2-b^2)\ {\rm with}\  \widetilde{A}(z)=O(1).$$
The leading correction to the large $z$ behaviour of $\omega_{\rm sc}$ is
found from equation \eompg\ to be of order $z^{-2/3}$. This determines the
singularity at $z=0$ and  implies $A(0)=B(0)$. The unique solution is thus
\eqn\eOmcr{\Omega_{\pm}=\pm{i\over \sqrt{3}z\sqrt{1-b^2/z^2}}
\left(\sqrt{1-b^2/z^2}\pm ib/z\right)^{2/3}.}
Therefore
\eqn\eomcrx{\omega_{\pm}(z,g)=\omega_{\pm}\left(z,g_c\right)\mp x g_c
4i\sqrt{3}{1\over z\sqrt{1-b^2/z^2}} \left(\sqrt{1-b^2/z^2}\pm
ib/z\right)^{2/3}+O\left(x^2\right) . }
By identifying the large $z$ expansion of $\omega_{\rm sc}$ with the small
$z$ expansion of \eomcrx\ we find the relation between the parameter
$a$ and $x$. For example for $z=-i\lambda$ and $\lambda\to 0_+$
we derive from the expansion \eomcrx:
\eqna\eomcras
$$\eqalign{\omega_+(-i\lambda)&\sim-2^{1/3}(1-\sigma)b^{4/3}\lambda^{2/3}
(1+O(x)),\cr
\omega_-(-i\lambda)&\sim -xg_c 3^{1/2}2^{8/3}b^{-1/3}\lambda^{-2/3}, \cr}$$
while the scaling function for $\lambda\to+\infty$
$$\omega_{{\rm sc},-}=-\ud (1-\sigma)b^{4/3}a^{4/3}\lambda^{-2/3}
+O\left(\lambda^{-4/3}\right).$$
It follows
\eqn\escalpg{a\sim b\sqrt{2}\left(19/81-\sigma/9\over
1-\sigma\right)^{3/4}x^{3/4}.}
Note finally that, taking into account the phase factors, $\omega_{\rm sc}$
can then also be written
\eqn\eompgt{\omega_{\rm sc}(z) = 2^{-1/3}b^{4/3}(1-\sigma)
\left[\left(-z+\sqrt{z^2-a^2}\right)^{2/3}
+\left(- z-\sqrt{z^2-a^2}\right)^{2/3}\right].}
This expression can be conveniently parametrized setting
\eqn\eparphi{z=-a\cos\varphi\,,\quad
\omega_{\rm sc}(z)=2^{2/3}a^{2/3}b^{4/3}(1-\sigma)\cos(2\varphi/3).}
We have assumed $\varphi\in[0,\pi]$ for $z\in[-a,+a]$.\par
\medskip
{\it Interpretation: the resolvent in the $(p,3)$ model.}
We recall that pure gravity is a $(2,3)$ model in the CFT
classification, which can be constructed as the solution of the canonical
commutation relation $[P,Q]=1$ where $P,Q$ are two differential operators of
order $2,3$ respectively \ref\rD{M. R. Douglas, {\it Phys. Lett.} B238 (1990)
176.}. \par
In the corresponding one-matrix model the saddle
point equations have a natural interpretation in terms of the resolvent of
the operator of order two. However, as we show in section 4, the trace
of the resolvent of the other differential operator is also solution of an
algebraic equation, of third degree. Solving the equation
explicitly, we obtain an expression proportional to expression \eompgt.
Moreover we verify that the scaling relation \escalpg\ is consistent with the
behaviour expected for the specific heat $u(x)$ in pure gravity.\par
Our results are thus consistent with the hypothesis
that the matrix  $S$  becomes, in the scaling limit, the
differential operator of order 3, $\d^3-(3/4)\{u,\d\}$ of pure gravity.
\subsec{The critical Ising model}
The line of pure gravity reaches the Ising critical point
when three zeros of $\Delta$ coincide: $a=e=0$. This situation is found for
$\sigma=1$ and thus $c=(-1+2\sqrt{7})/27$, $g_c=5\times
2^{-2} 3^{-9/2}b^3= 10c^3$ or $b=6\sqrt{3}c$.\par
The function $\omega_+$ then becomes
\eqn\eomIscr{\omega_+=z^2\left[(1-b^2/z^2)^{1/2}+ib/z\right]^{2/3}.}
For $z$ small for example $z=-i\lambda$, $\lambda\to 0_+$, we find that
$\omega_-$ gives the leading contribution to the two first terms of small $x$
expansion:
\eqn\ellarge{\omega_-=-(2b)^{2/3}\lambda^{4/3}- xg_c
3^{1/2}2^{8/3}b^{-1/3}\lambda^{-2/3} +O\left(x^2\right).}
This implies that $\omega_{\pm}(z)$ in the scaling limit, $g-g_c$ small,
have the form
$$\omega_{{\rm sc},\pm}=\vartheta_0 \left(\sqrt{a^2-z^2}\mp iz\right)^{4/3}.$$
Again $\vartheta_0$ and $a$ can be determined by expanding the scaling form
for $z$ large $z=-i\lambda$, $\lambda\to +\infty$:
$$\omega_{{\rm sc},-}(z) =\vartheta_0 (2\lambda)^{4/3}\left(1+{a^2\over
3\lambda^2}\right)+O\left(\lambda^{-4/3}\right).$$
Comparing with the expansion \ellarge\ we conclude
\eqna\escalIs
$$\eqalignno{\vartheta_0&=-\left(b/2\right)^{2/3} ,&\escalIs{a}\cr
a/b&=5^{1/2}3^{-3/2} x^{1/2}.&\escalIs{b}\cr}$$
As we show in section 4 these results agree, up to the normalizations of $x$
and $\omega$, with the scaling of the resolvent of the differential operator
$P=\d^3-(3/2)\{\d,u(x)\}$ of the $(3,4)$ model in the spherical limit.
We thus find complete consistency with the established results of the
critical Ising model in the spherical limit.\par
Note that, here again, it is convenient to use the parametrization \eparphi:
$$\omega_{\rm sc}(z)=\vartheta_0 \left[\left(-z+\sqrt{z^2-a^2}\right)^{4/3}
+\left(-z-\sqrt{z^2-a^2}\right)^{4/3}\right].$$
and thus
\eqn\eparphii{z=-a\cos\varphi,\ \Rightarrow\ \omega_{\rm sc}(z)=-2^{1/3}
b^{2/3} a^{4/3} \cos\left(4\varphi/3\right).}
\subsec{The singular free energy}
We show now how one can explicitly calculate the singular part
of the free energy $F=\ln Z$ and thus verify its universal character, when
compared to other quantities.\par
We start from
\eqna\eFreepr
$$\eqalignno{g^2{\del F\over \del g}&=N \bigl<\tr V(S)\bigr> \sim
N^2 \int\d s\,\rho(s)V(s) & \cr
& ={N^2\over 2i\pi}\oint\d z\, \omega_0(z)V(z) \, , & \eFreepr{} \cr}$$
where the contour in the last integral encloses the cut of $\omega_0(z)$.\par
We then multiply by $g$ and  differentiate again
\eqna\efreeI
$$\eqalignno{{\del \over \del g}\left(g^3{\del F\over \del g}\right)&=N^2 g
f(g),&\efreeI{a} \cr f(g)&={1\over2i\pi g}\oint\d z\, V(z)\Omega(z)\,.
&\efreeI{b} \cr}$$
where we have introduced the function $\Omega(z)$ defined by equation \eOmdef.
Using then the decomposition \eOmpm\ into \efreeI{} we find that we can write
\eqn\efreeIb{f={1\over2i\pi g}
\oint\d z\,\left[\e^{-2i\pi/3}V(z)-\e^{2i\pi/3}V(-z)\right]\Omega_+(z).}
The singular part of the free energy is thus related to the singular part
of $\Omega(z)$ for $a$ small which we determine now.
\medskip
{\it Expansion of $\Omega(z)$ for $a^2$ small.}
We recall that $\Omega(z)$ satisfies the homogeneous equation \egenRH.
Moreover $\Omega(z)$  has a cut along $[a,b]$ and  behaves for
large argument $z$ like $1/z$. It has singularities of the form
$(a-z)^{-1/2}$ and $(z-b)^{1/2}$ at $z=\pm a,\pm b$. It can be directly
determined by solving an equation of the form \eomegiii
\eqn\eqOM{\Omega^3(z)-3\pp(z)\Omega(z)-2\qq(z)=0\,.}
The functions $\pp,\qq$ which appear here can have simple poles
at $z=a,b$. This implies that they can be parametrized as
$$\pp(z)={1\over3}{z^2+\tau^2 d \over (z^2-a^2)(z^2-b^2)},\qquad
\qq(z)={2\over3\sqrt{3}}{b\tau^3 \over (z^2-a^2)(z^2-b^2)}. $$
The discriminant $\Delta$ of the equation $\Delta=\qq^2-\pp^3$
has as a numerator a polynomial of degree six which must be a perfect square,
condition which determines the parameters $d$ and $\tau$. However, it is more
transparent to use a different method. \par
The critical function ($a=0$) is given by equation \eOmcr:
$$\Omega_{\pm}=\pm{i\over \sqrt{3}z\sqrt{1-b^2/z^2}}
\left(\sqrt{1-b^2/z^2}\pm ib/z\right)^{2/3}.$$
The most singular, for $z$ small, of the two functions behaves like
$z^{-2/3}$.
This determines the corresponding scaling function. Indeed it has the general
form \eomscgen, $C$ can have poles at $z=\pm a$, and $C,D$ behave like $1/z^2$
for $z$ large. Thus
$$\Omega_{{\rm sc},\pm}={\vartheta_0 \over \sqrt{a^2-z^2}}
\left(\sqrt{a^2-z^2}\mp iz\right)^{1/3} ,$$
with
$$\vartheta_0=2^{1/3}3^{-1/2}b^{-1/3}.$$
Expanding $\Omega_{\rm sc}$ at next order for $z$ large we get a correction of
order $z^{-4/3}$. More precisely for $\lambda\to+\infty$
$$\Omega_{{\rm sc},+}(-i\lambda)\sim 3^{-1/2}a^{2/3}b^{-1/3}\lambda^{-4/3}.$$
This next to leading contribution must also be the small $z$ behaviour of the
correction to the critical function we are looking for. This correction on
the other hand is at most of order $1/z^2$ for $z$ large. These conditions
determine it entirely and we get
\eqn\edelOM{\Omega_{+}(z,a)-\Omega_{+}(z,0)\sim - (a/b)^{2/3}
{2^{-1/3}3^{-1/2} b\over z^2 \sqrt{1-b^2/z^2}}
\left(\sqrt{1-b^2/z^2}+ ib/z\right)^{-1/3}.}
Note that the expression in the r.h.s.\ has a singularity at $z=b$ only of
strength $(z-b)^{-1/2}$ while $(z-b)^{-3/2}$ could have been expected.
Actually it can be verified that the location of the singularity at $z=b$
varies by a term of order $a^2$ at least and thus  the corresponding
contribution to $\Omega$ is even smaller.
\medskip
{\it The free energy.}
We note the identity
$${\d \over \d z}\left(\sqrt{1-b^2/z^2}+ib/z\right)^{-1/3}
={i b\over 3}{1 \over  z^2 \sqrt{1-b^2/z^2}}
\left(\sqrt{1-b^2/z^2}+ib/z\right)^{-1/3}.$$
Introducing the expansion \edelOM\ into equation \efreeIb, using this identity
and integrating by parts we find
$$\eqalign{f(a)&=f(0)-{i 2^{-1/3}3^{1/2}
(a/b)^{2/3}\over 2i\pi g}\oint\d
z\,\left[\e^{-2i\pi/3}V'(z)+\e^{2i\pi/3}V'(-z)\right]
\cr&\quad\times \left(\sqrt{1-b^2/z^2}+ib/z\right)^{-1/3}.
\cr}$$
We now use the identity
$$\e^{-2i\pi/3}V'(z)+\e^{2i\pi/3}V'(-z)
=ig\sqrt{3}\left[\e^{2i\pi/3}\omega_0(z)-\e^{-2i\pi/3}\omega_0(-z)\right]
-{1 \over 4}\omega_-\,,$$
consequence of the definitions \esolo,\eomegdef, in the integral.\par
We need $\omega_-$ only at leading order. From \eomIscr\ we deduce
$$\omega_-=z^2 \left(\sqrt{1-b^2/z^2}+ib/z\right)^{-2/3}.$$
We then see that $\omega_-$ gives no contribution to the integral. We
calculate the
contribution of $\omega_0(z)$ by taking the residue at infinity.
We need only the large $z$ behaviour of $\omega_0(z)$ at leading order.
$$ig\sqrt{3}\left[\e^{2i\pi/3}\omega_0(z)-\e^{-2i\pi/3}\omega_0(-z)\right]
\sim -{i g\sqrt{3} \over  z}+O\left(z^{-2}\right).$$
We thus obtain
$$\oint{\d z\over
2i\pi g}\left[\e^{-2i\pi/3}V'(z)+\e^{2i\pi/3}V'(-z)\right]
\left(\sqrt{1-b^2/z^2}+ib/z\right)^{-1/3}=-i\sqrt{3}\,. $$
This completes the calculation of second derivative of the singular free
energy $F_{\rm sg}$ and we find:
\eqn\eFsg{g_c^2 F_{\rm sg}''(g)\sim F''(x)\sim{N^2\over 2i\pi g_c}\oint\d z\,
V(z)\Omega_{\rm sg}=-3 \times 2^{-1/3} N^2(a/b)^{2/3},}
with $x=1-g/g_c$. In the case of the $(3,2)$ ordinary critical point
this yields
$$F''(x)=-3 (19/81-\sigma/9)^{1/2}(1-\sigma)^{-1/2} N^2 x^{1/2}.$$
In particular for $c=0$ we recover the expected result, i.e.\ twice the value
of the corresponding one-matrix model.\par
For the critical Ising model instead we find
$$F''(x)=-(5/2)^{1/3}N^2 x^{1/3}.$$
\subsec{Large order behaviour}
We now look  for a non minimal saddle point of the integrand \eZIiii.
We know from our experience of the one-matrix model, that the second lowest
saddle point is obtained when we move only
one eigenvalue from the edge of the distribution $\rho$, to the nearest
stationary position. The variation of the action $\Sigma$ is then:
$$ \delta \Sigma= \int_{a}^{\lambda_f} \d \lambda {\partial \Sigma\over
\partial \lambda}, $$
where $\Sigma$ is given by
$$ {\partial \Sigma\over \partial \lambda}=N\left({V'(\lambda )\over
g}-{2\over N}\sum_j {1\over \lambda -\lambda _j}+{1\over N}\sum_j{1\over
\lambda +\lambda _j}\right).$$
In the continuum limit
\eqna\einstan
$$\eqalignno{{\partial \Sigma\over \partial \lambda}&=N\left({V'(\lambda
)\over g}-2\omega_0(\lambda ) -\omega_0(-\lambda)\right)&\cr
&=-{N\over12g}\bigl(2\omega(\lambda)+\omega(-\lambda)\bigr).
& \einstan{} \cr}$$
We are looking for a zero of $\partial \Sigma/\partial \lambda$ in the
interval $(0,a)$ and thus $\lambda_f$ is small. We can thus replace
$\omega(z)$  by its scaling form. We then find the variation $\delta F$ of
the free energy in the scaling limit. For the generic critical point, using
\eparphi\ and \escalpg\ we find
$$\delta \Sigma\propto a^{5/3}\propto x^{5/4},\ \Rightarrow\ \delta F\propto
\exp\left(-{\rm const.}\ x^{5/4}\right).$$
in agreement with the result found in pure gravity. For the Ising critical
point, using \eparphii\ and \escalIs{b} we obtain
$$\delta \Sigma\propto a^{7/3}\propto x^{7/6}, \ \Rightarrow\ \delta F\propto
\exp\left(-{\rm const.}\ x^{7/6}\right).$$
in agreement with the known result for the Ising model.
\subsec{Multicritical points}
With a general potential $V(S)$ higher order critical points
can be generated. If the polynomials $A,B,$ in expression \eomcrgen\
behave like: $A(z)\propto B(z)\propto z^{2m+2}$ for $z$ small, $\omega(z)
\propto z^{(6m+2)/3}$. The minimal example corresponds to
$$\eqalign{\omega_{\pm}&=\mp i (z/b)^{2m+1}
\left(\sqrt{1-b^2/z^2}\pm ib/z\right)^{-1/3},\cr
\pp(z)&=(z/b)^{4m+2},\quad \qq(z)=(z/b)^{6m+2}.\cr}$$
If in addition
$A(z)/B(z)\to 1$ then $\omega(z)\propto z^{(6m+4)/3}$. The minimal model is
$$\eqalign{\omega_{\pm}&=(z/b)^{2m+2}
\left(\sqrt{1-b^2/z^2}\pm ib/z\right)^{2/3},\cr
\pp(z)&=(z/b)^{4m+4},\quad \qq(z)=(z/b)^{6m+6}-2(z/b)^{6m+4}.\cr} $$
(Note the change of normalization compared to the $m=0$ case.)
\medskip
{\it Scaling region.}
It will be convenient in what follows to classify the critical models
when $\omega(z)$ behaves for $z\to 0$ like
$$\omega(z)\propto z^{p/q},$$
in terms of two relatively prime integers $p,q$, in analogy with the CFT
classification (see section 4). For all the models considered up to now
$q=3$, while $p$ belongs to one of the two sets $p=6m+2,6m+4$.\par
To determine the  functions in the scaling limit in terms of the deviation
$x$ from the critical cosmological constant we now use the same method as in
the case of the simple critical points $m=0$. We first expand $\omega(z)$ at
first order in $x$ and then look for the leading term for $z\to 0$.
The two cases have to be discussed separately.
\medskip
{\it First case $p=6m+2$.}
At leading order for $z=-i\lambda$, $\lambda\to 0_+$ we find
$$\omega_+(z)\sim(-1)^{m+1}2^{1/3}(\lambda/b)^{2m+2/3}.$$
Therefore the scaling function $\omega_{\rm sc}$ has the form \eomscgen\ with
$C,D$ polynomials of degree $2m$.\par
It is convenient to normalize the potential in such a way that the variation
$\delta\omega$ at leading order in $x$ and for $z$ large is
$$\delta\omega_{\pm}\sim \mp ix b/z.$$
Then $\delta\omega_{\pm}$, which is independent of the potential and thus
proportional to $\Omega_{\pm}$ of equation \eOmcr, is
\eqn\edelom{\delta\omega_{\pm}=\mp{ ix b \over z\sqrt{1-b^2/z^2}}
\left(\sqrt{1-b^2/z^2}\pm ib/z\right)^{2/3}.}
Expanding $\delta\omega_{-}$ for $\lambda\to 0_+$  we find
$$\delta\omega_-(-i\lambda)\sim-x 2^{2/3}b^{2/3}\lambda^{-2/3}.$$
This behaviour of $\delta\omega$ determines completely the scaling function.
It can be expressed in terms of the function $\vartheta(z)$:
\eqn\eVintIs{\vartheta(z)= \int_{-iz}^{\sqrt{a^2-z^2}}\d
t\,(t+iz)^{m-1/3}(t-iz)^m\,.}
Calculating the integral we first verify that the function
$\vartheta(z)$ has a form  \eomscgen:
$$\vartheta(z)=\left(\sqrt{a^2-z^2}+ iz\right)^{-1/3}\left(C(z)
\sqrt{a^2-z^2}+ iz D(z)\right).$$
For $\lambda\to +\infty$ we find
$$\eqalign{\vartheta(-i\lambda)&\sim(-1)^m B(m+1,m+2/3) (2\lambda)^{2m+2/3},
\cr \vartheta(i\lambda)&\sim
{a^{2m+4/3}(2\lambda)^{-2/3} \over m+2/3}, \cr}$$
with
$$B(\alpha,\beta)={\Gamma(\alpha)\Gamma(\beta)\over \Gamma(\alpha+\beta)}.$$
It follows
\eqn\eomscg{\omega_{{\rm sc},\pm}(z)=\vartheta_0 \vartheta(\pm z),}
with
$$\eqalign{\vartheta_0&=-{2^{-2m-1/3}b^{-2m-2/3}
\over B(m+1,m+2/3)},\cr
a/b&=2\left[2^{1/3}(m+2/3)B(m+1,m+2/3)x\right]^{1/(2m+4/3)} .\cr}$$
Using equation \eFsg\ we conclude that the singular part of the free energy
then scales like
$$F''(x)\propto x^{1/(3m+2)}.$$
As we show in section 4 all these results are consistent with the behaviour
found in the $(6m+2,3)$ string model. Moreover equation \einstan{} yields
then the expected scaling for the ``instanton" action $\delta \Sigma\propto
x^{1+1/(6m+4)}$.
\medskip
{\it Case $p=6m+4$.} In the $\lambda\to 0_+$ limit $\omega_-$
gives the leading contributions:
$$\omega_-=(-1)^{m+1}2^{2/3}(\lambda/b)^{2m+4/3}-x2^{2/3}(b/\lambda)^{2/3}
+ O\left(\lambda^{-4/3}\right).$$
We introduce now the function $\vartheta(z)$:
\eqn\eVintIsb{\vartheta(z)= \int_{-iz}^{\sqrt{a^2-z^2}}\d
t\,(t+iz)^{m+1/3}(t-iz)^m\,.}
For $\lambda\to +\infty$ we find:
$$\vartheta(-i\lambda) =(-1)^m B(m+1,m+4/3)(2\lambda)^{2m+4/3}
+{2^{-2/3}a^{2m+2}\over m+1}\lambda^{-2/3}
+O\left(\lambda^{-4/3}\right).$$
In the scaling limit we conclude that $\omega(z)$ must have the form
\eomscg\ with now
$$\omega_{\pm}(z)=\vartheta_0\vartheta(\mp z),$$
with
$$\eqalign{\vartheta_0&=-{b^{-2m-4/3}2^{-2m-2/3}\over
B(m+1,m+4/3)},\cr
{a\over b}&=2\left[(m+1)B(m+1,m+4/3) x\right]^{1/(2m+2)}.\cr}$$
It follows that the specific heat $F''(x)\propto x^{1/(3m+3)}$.
Again these results are consistent with a $(6m+4,3)$ model (see section 4).
Agreement is also found with the scaling form of the
instanton action $\delta \Sigma\propto x^{1+1/(6m+6)}$.
\newsec{The general $O(n)$ model}
We now generalize the previous model by replacing the matrix $A$ by a set of
$n$ matrices $A_i$:
\eqn\eZOn{Z=\int \d S\,\d A_1\ldots \d A_n\, \ee{-(N/ g)\tr
[S(A^2_1+\cdots+ A^2_n)+V(S)]}, }
$V(S)$ being now also a general polynomial potential.\par
The model has then an $O(n)$ symmetry. The quantity
$F=\ln Z$ can be interpreted as the free energy of a gas of loops, each
indexed by an integer $i$, $i=1,...,n$,  drawn on a random lattice of the
form of a Feynman diagram.\par
The corresponding model on regular lattices can become critical only for
$-2\le n\le 2$. We shall verify that here that the integral is even defined
only for $n\le 2$. It is thus convenient to set $n=-2\cos{\theta}$. Although
we could restrict ourselves to the interval $0\le \theta
\le \pi$, it is convenient for book-keeping purpose, to consider all positive
values of $\theta$. Note that the case $n=0,\theta=(2m+1)\pi/2$ reduces to the
standard one-matrix model. For $n=1$ we can assign  $\theta=p\pi/3$
to the multicritical models when $\omega(z)\propto z^{p/3}$; the critical
Ising model thus corresponds to $\theta=4\pi/3$.
\par
The integral over the matrices $A_i$'s is still gaussian and can be performed.
We can then parametrize $S$ in terms of a unitary transformation and its
eigenvalues $\lambda_i$. After the integration over unitary transformations,
the integral \eZOn\ becomes:
$$\eqalignno{Z&=\int  \Delta^2(\Lambda)
\prod_{i,j}(\lambda_i+\lambda_j)^{-n/2}\prod_i \d\lambda_i
\ee{-(N/ g) V(\lambda_i)}&\cr
&=\int \d\lambda\, \ee{-N\, \Sigma[\lambda]},&\cr} $$
with the effective action $\Sigma$
$$ \Sigma =\sum_i {1\over g}V(\lambda_i)-{1\over N}\sum_{i\neq j}
\ln{\vert\lambda_i-\lambda_j\vert}+{n\over 2N}\sum_{i,j}
\ln(\lambda_i+\lambda_j). $$
\subsec{The saddle-point equation}
In the planar limit $N\to\infty$, $Z$ can be calculated by the steepest
descent method. The saddle point equation is:
\eqn\eesaddle{{\partial \Sigma\over\partial\lambda_i}=0={1\over
g}V'(\lambda_i)-{2\over N}\sum_{j\neq i} {1\over\lambda_i-\lambda_j}+{n\over
N}\sum_j{1\over\lambda_i+\lambda_j}. }
We again introduce the density of eigenvalues $\rho(\lambda)=
{1\over N}\sum_i \delta(\lambda-\lambda_i)$, and its Hilbert's transform, the
trace of the resolvent:
$$ \omega_0(z)={1\over N}\sum_i {1\over z-\lambda_i}=\int \d\lambda
{\rho(\lambda)\over z-\lambda}. $$
In the large $N$ limit $\rho(\lambda)$ becomes a continuous function and
$\omega_0$ a function analytic  except when $z$ belongs to the spectrum of
$S$, that is with a cut on a segment $[a,b]$ of the real
positive axis.\par
Equation \eesaddle\  may be written in term of $\omega_0$:
\eqn\eqwze{{\omega_0(\lambda+i0)+\omega_0(\lambda-i0)+n\omega_0(-\lambda)
={1\over g}V'(\lambda),\qquad (\lambda\in [a,b])}.}
This linear equation has a polynomial solution:
\eqn\eqwr{\omega_r(z)={1\over g}{1\over 4-n^2}\bigl(2V'(z)-nV'(-z)\bigr).}
Note that the cases $n=\pm 2$ are special and must be examined separately.
The function $\omega(z)$, defined by
\eqn\eomreg{\omega_0=\omega_r+\omega/g,}
then satisfies the homogeneous equation:
\eqn\eqw{\omega(\lambda+i0)+\omega(\lambda-i0)+n\omega(-\lambda)=0\,.}
Since $\omega_0(z)$ behaves as $1/ z$ for $z$ large, $\omega(z)$ has the
large $z$ expansion:
\eqn\ezlarge{\omega(z)
=-{4\over \sqrt{2-n}}\bigl(2V'(z)-nV'(-z)\bigr) +{g \over z}
+O\left(z^{-2}\right)\,. }
\medskip
{\it A quadratic relation.} Let us introduce the following function:
\eqn\edefr{\rr(z)=\omega^2(z)+\omega^2(-z)+n\omega(z)\omega(-z).}
We verify, as in the case of the Ising model, that the
discontinuity on the cut of $\rr(z)$ vanishes as a consequence of equation
\eqw:
$$\eqalign{\rr(z+i0)-\rr(z-i0)&=\left[\omega(z+i0)-\omega(z-i0)\right] \cr
&\quad \times\left[\omega(z+i0)+\omega(z-i0)+n\omega(-z)\right]=0\,.}$$
Therefore $\rr$ is an even function, analytic  in the whole complex plane. The
behaviour of $\omega$ for $z$ large, then implies that $\rr$ is a polynomial.
Again equation \edefr\ can be directly derived from the saddle
point equation or the loop equation. An expression of $\rr(z)$ in terms of
the potential follows.
\subsec{The general solution}
We introduce two auxiliary functions:
\eqn\edefwpm{\left\lbrace\eqalign{
\omega_{+}(z)&={i \over 2\sin\theta}\left[\e^{i\theta/2}\omega(z)-
\e^{-i\theta/2}\omega(-z)\right]\cr
\omega_{-}(z)&=-{i \over 2\sin\theta}
\left[\e^{-i\theta/2}\omega(z)-\e^{i\theta/2}\omega(-z)\right]  \cr}
\right.}
such that $\omega_{+}(-z)=\omega_{-}(z)$. Then
$$\omega_{+}(z)\omega_{-}(z)=\pp(z).$$
Conversely $\omega(z)$ is given in terms of $\omega_+,\omega_-$  by
\eqn\esolw{\omega(z)=-\left[\e^{i\theta/2}\omega_{+}(z)
+\e^{-i\theta/2}\omega_{-}(z)\right].}
Equation \eqw\ then is equivalent to the simple relations
\eqn\edisc{\omega_{\pm}(z-i0)=\e^{\pm i\theta}\omega_{\mp}(z+i0), }
which, as in Ising case, imply that $\omega(z)$ has cuts only on the positive
axis. \par
If we now choose $\theta$ such that $ \e^{iq\theta}=\pm 1$, $q$ being a
positive integer, then $\omega(z)$ satisfies an algebraic equation of degree
$q$. We have to examine the two signs separately.
\medskip
{\it Case $\e^{iq\theta}=1$.} This implies $\theta=\pi p/q$,
where $p$ is an even integer. The function $\qq(z)$,
\eqn\edefq{\qq(z)=\ud (\omega_{+}^q+\omega_{-}^q), }
has no discontinuity on the cut, and is analytic in the whole complex
plane. It is therefore an even polynomial of a degree determined by the
degree of the potential .
We have thus the two following algebraic equations:
\eqn\eompqk{\omega_{+}\omega_{-}=\pp(z),\qquad
\omega_{+}^q+\omega_{-}^q =2\qq(z)\,. }
The solution of these equations is then $\omega_{\pm}^q=\qq\pm\sqrt{\Delta}$,
with
\eqn\edisq{\Delta=\qq^2-\pp^q\,,\qquad
\sqrt{\Delta}=\ud\left(\omega^q_+-\omega^q_-\right). }
Note that $\sqrt{\Delta}$ is thus an odd function.
The function $\omega$ is then given by equation \esolw.
\medskip
{\it Case $\e^{iq\theta}=-1$.}  This implies $\theta=\pi p/q$,
where $p$ is now an odd integer. The expressions are quite similar but the
role of $\qq(z)$ and $\sqrt{\Delta}$ are formally exchanged. It is now the
function $\qq(z)$,
\eqn\edefqm{\qq(z)={1\over2i}(\omega_{+}^q-\omega_{-}^q),}
which has no discontinuity on the cut, and is analytic in the whole complex
plane. It is therefore an odd polynomial.
We have thus the two following algebraic equations:
\eqn\eompqkm{\omega_{+}\omega_{-}=\pp(z), \qquad
\omega_{+}^q-\omega_{-}^q=2i\qq(z)\,. }
The solution of these equations is $\omega_{\pm}^q=\sqrt{\Delta}\pm i\qq$,
with
$$\Delta=\pp^q-\qq^2\,,\qquad
\sqrt{\Delta}=\ud\left(\omega^q_+ +\omega^q_-\right). $$
Here instead $\sqrt{\Delta}$ is even. The function $\omega$ is still  given
by equation \esolw.
\medskip
{\it One-cut solution.}
We still have to determine the coefficients of the polynomials
$\pp,\qq,\Delta$. They can be found from the additional condition that
$\omega$ has only one cut $[a,b]$ on the positive real axis.
We can see from \esolw\ that the singularities of $\omega$ are the single
roots of $\Delta$. We demand  that
except for $a$ and $b$ (and $-a$,$-b$ by parity), all the roots of $\Delta$
are double, in such a way that  $\Delta$ can be written:
$$ \Delta=-(z^2-a^2)(z^2-b^2)M^2(z),$$
where $M(z)$ is an odd or even polynomial depending on the different cases.
Due to the special form of the conditions \edisc\ all one-cut solutions
$\omega_{\pm}$ can be factorized:
$$\omega_{\pm}=\Omega_{\pm}(z)\left[\pm z A(z)+
B(z)\sqrt{(z^2-a^2)(z^2-b^2)}\right] ,$$
where $A$ and $B$ are even functions, rational fractions in general because
$\Omega{\pm}(z)$ may have zeros, and the function $\Omega$, which has
only singularities at $\pm a,\pm b$, is a ``minimal" solution of:
$$\Omega_+(z)=\Omega_-(-z),\qquad
\Omega_{\pm}(z-i0)=-\e^{\pm i\theta}\Omega_{\mp}(z+i0).$$
This factorization property is a consequence of the algebraic equation
satisfied by $\omega(z)$ and may not necessarily hold when $\theta/\pi$ is
not rational.
\subsec{Critical points}
We again consider only critical points for which $a=0$. The function
$\omega(z)$ at a critical point has then a cut for $0\le z\le b$. The general
form of such a solution is analogous to the form \eomcrgen\ of the case $n=1$:
\eqn\eompqcrg{\omega_{\pm}(z)=\left(\sqrt{1-b^2/z^2}\pm ib/z\right)^{-l/q}
\left[A(z)\sqrt{1-b^2/z^2} \pm i b B(z)/z\right],}
where $l,q$ are relatively prime integers with $0<l<q$ and
$A, B$ are polynomials which can be chosen even without loss of generality.
Indeed the situation $A,B$ odd is equivalent to $A,B$ even with the
change $l\mapsto q-l$.\par
One immediately  verifies
that $\pp(z),\qq(z)$ and $\Delta(z)$ are polynomials of a form consistent
with a one-cut solution provided $A$ and $B$ vanish at $z=0$. \par
A minimal realization of a critical point with polynomial potentials is:
\eqna\ecrpq
$$\eqalignno{\omega_{\pm}(z)&=\mp i(z/b)^{2m+1} \left(\sqrt{1-b^2/z^2}\pm
ib/z\right)^{-l/q},& \ecrpq{a} \cr
\omega_{\pm}(z)&=(z/b)^{2m+2} \left(\sqrt{1-b^2/z^2}\pm
ib/z\right)^{1-l/q}.& \ecrpq{b} \cr} $$
Then in both cases
\eqn\ethetag{\omega_+(z-i0)=\e^{i\pi(1-l/q)}\omega_-(z+i0)\ \Rightarrow\
\theta=\pi(1-l/q) .}
The two cases $q-l$ even and odd  correspond to
the two situations $\e^{iq\theta}=\pm 1$.\par
For $z\to 0$ we find that
$$\omega(z)\propto z^{p/q}$$
where $p=(2m+1)q-l$ in the case $(a)$ and $p=(2m+1)q+l$ in the case $(b)$.
Note that the values of $p$ are such that $m$ can also be defined as the
integer part of $p/2q$ since $p/(2q)-1<m<p/(2q)$.
Finally we see that for book-keeping purpose it is convenient to assign the
angle $\theta=\pi p/q$ to the critical point characterized by the integers
$(p,q)$.
\subsec{Scaling region}
We want now to derive $\omega(z)$ in the scaling region when the variable
$x=1-g/g_c$ which characterizes the deviation of the cosmological constant
from its critical value is small. Functions which satisfy equation \eqw\
and are singular only at $z=\pm a$ and $z=\infty$ have the general form:
\eqn\eomscgen{\omega_{{\rm sc,}\pm}
=\left(\sqrt{a^2-z^2}\pm iz\right)^{-l/q}\left[C(z)
\sqrt{a^2-z^2}\pm iz D(z)\right],}
where again $C$ and $D$ are even polynomials.
A comparison between the large $z$ behaviour of $\omega_{\rm sc}$ and the
small $z$ behaviour of $\omega$ at the critical point  yields the degrees of
$C$ and $D$.
This determines them completely only for the minimal critical points $m=0$.
We then find
$$\omega_{{\rm sc,}\pm}\propto \left(\sqrt{a^2-z^2}\pm
iz\right)^{p/q}.$$
To obtain the relation between $x$ and $a$ and completely
determine the form of the polynomials $C,D$ for multicritical points ($m>0$)
we then proceed as in the $q=3$ case and
calculate the deviation from the critical form at leading order for $x$ small.
\medskip
{\it Deviation from the critical form at leading order.}
We now calculate the deviation from the critical form at leading order
in the variable $x=1-g/g_c$ which characterizes the deviation of the
cosmological constant from its critical value. We normalize the potential in
such a way that for $z$ large the variation $\delta\omega_{\pm}$ is:
$$\delta\omega_{\pm}\sim \mp ixb/z\,.$$
As in the case of the Ising model we introduce the function
$\Omega(z)$:
$$\Omega(z)={\partial\over\partial g}\bigl(g\omega_0(z)\bigr).$$
Equation \eomreg\ then implies
\eqn\eOmdefg{\Omega(z)= {\partial\omega(z)\over\partial g}.}
It thus satisfies the homogeneous equation \eqw. We also introduce the
decomposition
\eqna\eOmpmg
$$\eqalignno{
\Omega_{\pm}(z)&=\pm{i \over 2\sin\theta}\left[\e^{\pm i\theta/2}\Omega(z)-
\e^{\mp i\theta/2}\Omega(-z)\right]&\eOmpmg{a}\cr
\Omega(z)&=-\left[\e^{i\theta/2}\Omega_+(z)
+\e^{-i\theta/2}\Omega_-(z)\right].&\eOmpmg{b}\cr}$$
{}From the definition of $\omega_0$ we infer the behaviour of $\Omega$ for
$z$ large, $\Omega(z)\sim 1/z$. Moreover, since $\Omega$ is the derivative
of a function which has singularities of the form $(z-z_0)^{1/2}$, $z_0=\pm
a,\pm b$, it can have
a stronger singularity of $(z-z_0)^{-1/2}$ type. These conditions determine
$\Omega(z)$ uniquely as a function of the location of the singularities, as
in the Ising case. The variation $\delta\omega$ is proportional to $\Omega$
at the critical point. At the critical point $\Omega$ has the form \eompqcrg.
To obtain its complete form we need its small $z$ behaviour. This behaviour
must be consistent with the leading correction to the large $z$ behaviour of
$\omega_{\rm sc}$. Since $\Omega$ is independent of the potential we can
compare it to the form of $\omega_{\rm sc}$ in the case $(a)$ for $m=0$.
The leading correction to $\omega_{\rm sc}$ is then of order $z^{l/q-1}$.
The unique solution is then
$$\delta\omega_{\pm}=\mp {ixb \over z\sqrt{1-b^2/z^2}}
\left(\sqrt{1-b^2/z^2}\pm ib/z\right)^{1-l/q}.$$
With this information we can now explicitly calculate the
scaling functions for all critical points.
\medskip
{\it Scaling function.} From the preceding analysis we conclude that for $z$
large $\omega_{\rm sc}$ satisfies
\eqn\econd{\omega_{\rm sc}(z)-{\rm const.\ }z^{p/q}
=O\left(x z^{l/q-1}\right).}
It is easy to verify that this fixes the polynomials $C,D$. We now show
that the scaling
function $\omega_{\rm sc}$ can then be expressed in terms of the function
$\vartheta(z)$ given by the integral representation
\eqn\eVint{\vartheta(z)= \int_{-iz}^{\sqrt{a^2-z^2}}\d
t\,(t+iz)^{p/q-m-1}(t-iz)^m\,,}
the proof relying on a verification of condition \econd.
Calculating the integral \eVint\ we first verify that the function
$\vartheta(z)$ has a form consistent with expression \eomscgen:
$$\vartheta(z)=\left(\sqrt{a^2-z^2}+ iz\right)^{\pm l/q}\left(C(z)
\sqrt{a^2-z^2}+ iz D(z)\right),$$
where $C,D$ are two even polynomials of degree $2m$ and $\pm l/q=p/q-2m-1$.
Moreover if we introduce the parametrization of case $(a)$, $p=(2m+1)q-l$,
we find
$$\vartheta(l,z)= \int_{-iz}^{\sqrt{a^2-z^2}}\d
t\,(t+iz)^{(p-q-l)/(2q)}(t-iz)^{(p-q+l)/(2q)}.$$
It follows
$$\vartheta(l,z)-\vartheta(-l,-z)=(2z)^{p/q}\e^{i\pi(q-l)/(2q)}\sigma_{pq},$$
where we have set
\eqn\esigma{\sigma_{pq}=B(m+1,p/q-m)={\Gamma[(p+q+l)/(2q)]\Gamma[(p+q-l)/(2q)]
\over \Gamma[(p+q)/q]} .}
We then expand $\vartheta(\pm z)$ for $z$ large.
\medskip
{\it Large $z$ expansion.} For $z=-i\lambda$ large we find
\eqna\eVzlarge
$$\eqalignno{\vartheta(-i\lambda)&= (-1)^m
(2\lambda)^{p/q}{\Gamma(m+1)\Gamma(p/q-m)
\over \Gamma(p/q+1)}+(2\lambda)^{p/q-2m-2}{a^{2m+2}\over
m+1} & \cr &\quad +O\left(\lambda^{p/q-2m-4}\right), &\eVzlarge{a}\cr
\vartheta(i\lambda) &\sim {(2\lambda)^{2m-p/q} a^{2p/q-2m} \over p/q-m}.
&\eVzlarge{b}\cr}$$
\medskip
{\it Case $(a)$.} If $p=(2m+1)q-l$, and thus $p/q<2m+1$, $\vartheta(-z)$ is
asymptotically larger than the correction to $\vartheta(z)$. Moreover
$2m-p/q=l/q-1$. The solution is then
\eqn\eomscpqa{\omega_{{\rm sc},\pm}(z)=\vartheta_0\vartheta(\pm z).}
Moreover comparing the expansion \eVzlarge{} with the expansions of
the critical functions $\omega_{\pm}$ and $\delta\omega_{\pm}$:
$$\eqalign{\omega_+(-i\lambda)&\sim(-1)^{m+1} 2^{l/q}(\lambda/b)^{p/q} ,\cr
\delta\omega_-(-i\lambda)&\sim - x 2^{1-l/q}(\lambda/b)^{l/q-1} ,\cr}$$
we obtain the normalization constant $\vartheta_0$ and the relation between
$a$ and $x$
$$\eqalign{\vartheta_0 &=-2^{(l-p)/q}b^{-p/q}\sigma_{pq}^{-1},\cr
\left(a\over b\right)^{(p+q-l)/q}&= 2^{(p+q-3l)/q}[(p+q-l)/q]\sigma_{pq} x,}$$
where we have used the definition \esigma.
\medskip
{\it Case $(b)$}. If $p=(2m+1)q+l$, and thus $p/q>2m+1$, the correction to
$\vartheta(z)$ is asymptotically larger than the correction to
$\vartheta(-z)$. Moreover $p/q-2m-2=l/q-1$.
The solution is then
\eqn\eomscpqb{\omega_{{\rm sc},\pm}(z)=\vartheta_0\vartheta(\mp z),}
with
$$\eqalign{\vartheta_0 &=- 2^{(q-l-p)/q}b^{-p/q}\sigma_{pq}^{-1}, \cr
\left(a \over b\right)^{(p+q-l)/q}&= 2^{(p-l)/q}[(p+q-l)/q] \sigma_{pq} x\,.
\cr}$$
Note that in the set of variables $p,q,l$ the behaviour of $a$ takes the same
form in both cases.
\subsec{The singular free energy}
We can find the singular part of the free energy, using the same method
as in the Ising model case. We have shown that
\eqn\efreeder{ {\partial\over \partial g}\left(g^3{\partial F\over \partial
g}\right) ={N^2\over 2i\pi}\oint V(z)\Om(z)\, \d z\, ,}
where $\Omega(z)$ is the function \eOmdefg. Using the decomposition
\eOmpmg{b} we can rewrite equation \efreeder
\eqn\eFrb{{\partial\over \partial g}\left(g^3{\partial F\over \partial
g}\right) =-{N^2\over 2i\pi} \oint \Om_{+}(z)\left(\e^{i\theta
/2}V(z)-\e^{-i\theta /2}V(-z)\right) \d z\,.}
The critical function for $a=0$ is
\eqn\eOmcrq{\Om_{\pm}=\pm{i\over 2\sin(\theta/2) z\sqrt{1-{b^2/ z^2}}}
\left(\sqrt{1-{b^2/ z^2}} \pm i{b/ z}\right)^{1-l/q}.}
\medskip
{\it The scaling region.}
Let us now consider the case $a\neq 0$ but small. For $z$ small the function
\eOmcrq\ behaves like $z^{l/q-1}$. This, together with the other properties,
determines the scaling form of $\Omega(z)$
$$\Omega_{{\rm sc},\pm}(z)={\vartheta_0\over\sqrt{a^2-z^2}}
\left(\sqrt{a^2-z^2}\mp iz\right)^{l/q} , \qquad
\vartheta_0={2^{-2l/q}b^{-l/q} \over \sin(\theta/2)}. $$
Conversely, as in the $q=3$ case, the next to leading term in the large $z$
expansion of $\Omega$ provides the additional information needed to completely
determine the first correction to the critical function for $a$ small. This
correction behaves like $z^{-1-l/q}$, therefore
$$\Omega_+(a)-\Omega_+(a=0)\propto {1 \over z^2 \sqrt{1-b^2/z^2}}
\left(\sqrt{1-b^2/z^2}+ ib/z\right)^{-l/q}.$$
The leading correction to $\Omega_{{\rm sc},+}(z)$ is:
$$\Omega_{{\rm sc},+}(-i\lambda)\sim \vartheta_0 2^{-l/q}a^{2l/q}
\lambda^{-1-l/q} .$$
We thus have:
$$\Om_{+}(a)-\Omega_{+}(0)\sim -{2^{-4l/q}\over \sin(\theta/2)}\left({a \over
b}\right)^{2l/q}  {b \over z^2 \sqrt{1-b^2/z^2}}
\left(\sqrt{1-b^2/z^2}+ ib/z\right)^{-l/q}.$$
The identity
$${\d \over \d z} \left(\sqrt{1-b^2/z^2}+ ib/z\right)^{-l/q}=
ib{l\over q} {1 \over z^2 \sqrt{1-b^2/z^2}}\left(\sqrt{1-b^2/z^2}+
ib/z\right)^{-l/q} ,$$
allows to cast this expression into the form
$$\Om_{+}(a)-\Om_{+}(0)=-i{q \over l\sin(\theta/2)} \left({a\over
4b}\right)^{2l/q}{\d \over\d z}\left(\sqrt{1-{b^2/
z^2}}+i{b/ z}\right)^{-l/ q}.$$
With this expression we can integrate by parts integral \eFrb. The second
derivative of the singular part of the free energy is then given by
$$\eqalignno{g_c^2 F''_{\rm sg}&={N^2\over 2i\pi g_c}{i q\over l\sin{\theta
/2}}\left({a \over 4b}\right)^{2l/q} \oint\d z \left(\sqrt{1-{b^2/
z^2}}+i{b/ z}\right)^{-l/ q} &\cr &\quad \times \left(
\e^{i\theta/2}V'(z)+\e^{-i\theta/2}V'(-z)\right).&\cr}$$
We then use the identity
$$\eqalign{\e^{i\theta/2}V'(z)+\e^{-i\theta/2}V'(-z)&
=2ig\sin\theta\left(\e^{-i\theta/2}\,\omega_0(z)
-\e^{i\theta/2}\omega_0(-z)\right)\cr&\quad +4\sin^2\theta\,
\omega_-(z).\cr}$$
We need $\omega_-$ only at leading order as given by expressions \ecrpq{}.
We see that the contribution to the integral coming from $\omega_-$ vanishes.
The contribution due to $\omega_0$ can be calculated by taking the residue at
infinity. Then only the leading behaviour of $\omega_0$ for $z$ large is
relevant:
$$2ig\sin\theta\left(\e^{-i\theta/2}\omega_0(z)
-\e^{i\theta/2}\omega_0(-z)\right)\sim {4ig \sin\theta\cos(\theta/2)\over
z}.$$
In terms of the variable $x=1-g/g_c$ we finally obtain:
$$g_c^2 {\d^2 F_{\rm sg} \over (\d g)^2}=F''_{\rm sg}(x)
=-N^2 (q/l)(2-n) 2^{1-4l/q}\left({a\over b}\right)^{2l/q}.$$
If we set $q=3$ and $l=1$ we find $3\times 2^{-1/3} \left({a/
b}\right)^{2/3}$, in agreement with the Ising model result.\par
We derive from this expression the scaling of the free energy for all
critical points:
\eqn\eFrscpq{F''_{\rm sg}(x)\propto x^{2l/(p+q-l)},\ \Rightarrow\ \gamma_{\rm
string}= -{2l\over p+q-l}.}
\subsec{Large order behaviour}
The calculation of section 2.7 can easily be generalized.
The variation of the action is:
$$ \delta \Sigma= \int_{a}^{\lambda_f} \d\lambda {\partial \Sigma\over
\partial\lambda} $$
and we have:
$$\eqalignno{ {\partial \Sigma\over
\partial\lambda}&=N\left({V'(\lambda)\over g} -{2\over N}\sum_j
{1\over\lambda-\lambda_j} +{n\over N}\sum_j{1\over\lambda+\lambda_j}\right),
&\cr
&=N\left({V'(\lambda)\over g}-2\omega_0(\lambda)-n
\omega_0(-\lambda)\right)&\cr
&=-N \left[2\omega(\lambda)+n\omega(-\lambda) \right]/g.&\cr}$$
In the critical region $g$ close to $g_c$, the argument $\lambda$ in the
integral remains of order $a$. We conclude that $\delta \Sigma$ scales like
$$\delta \Sigma\propto N x^{(p+q)/(p+q-l)},$$
result consistent with the scaling of the free energy. Thus the variation
$\delta F$ of the free energy has the form
\eqn\edelFsc{\delta F\propto \exp\left(-{\rm const.}\
x^{(p+q)/(p+q-l)}\right),}
result consistent with the $2k!$ behaviour at large orders of the topological
expansion also found in other matrix models.
\newsec{The resolvent in $(p,q)$ string models}
We have solved the saddle point equations of the $O(n)$ model by transforming
them into an algebraic equation for the trace of the resolvent of some matrix
in the large $N$ limit. It is therefore interesting, for comparison purpose,
to discuss the form of the resolvent in multimatrix models. The string
equations of the  $(p,q)$ models can be generated by constructing a
representation of the
canonical commutation relations in terms of two differential operators $P,Q$
of order $p,q$ respectively ($p,q$ are relatively prime and not both odd):
$$\eqalign{P&=\d^p-(p/4)\{\d^{p-2},u(x)\}+\sum_{i=2}\{v_i(x),\d^{p-2i}\},\cr
Q&=\d^q-(q/4)\{\d^{q-2},u(x)\}+\sum_{i=2}\{w_i(x),\d^{q-2i}\},\cr}$$
where $\d$ means $\d/\d x$ and $\{ .,.\}$ means anticommutator. Moreover
$u(x)$ is the specific heat, the second derivative of the free energy.\par
Let us consider the trace of the resolvent of one of these
operators, for example $Q$:
$$\omega(z)=\tr(z-Q)^{-1}.$$
In the spherical, i.e.\ the semiclassical limit, because the non commutation
between space and derivative can be neglected, $\omega(z)$ is given by
\eqn\eompq{\omega(z,x)=-\int^x\d x'\int{\d y \over
(iy)^q-(q/2)u(x')(iy)^{q-2}+2\sum_{i=2}w_i(x')(iy)^{q-2i}-z}.}
The integral over $y$ just selects a residue.\par
The scaling relations in the $(p,q)$ model, in the same limit, are
$$w_i(x)\propto u^i(x),\qquad u(x)\sim x^{2/(p+q-1)}.$$
Thus $\omega(z,x)$  has the scaling form:
$$\omega(z,x)=z^{p/q}\omega\left(1,xz^{-(p+q-1)/q}\right).$$
Moreover in the critical limit, i.e.\ for $z$ large, it behaves like
$$\omega(z,x)\propto z^{p/q}+x z^{1/q-1}+o\left(z^{1/q-1}\right).$$
Note immediately that these scaling properties are consistent with those
of the $O(n)$ model studied above if and only if $l=1$, i.e.\ $p$ has the
special form $p=(2m+1)q\pm 1$.\par
We have shown that in the case of  the $O(n)$ models $\omega(z)$ satisfies,
in the large $N$ limit, an algebraic equation of a special type.
We show now that in this special class of $(p,q)$ models the resolvent
satisfies the same equations. \par
It is well known that in the case of the one-matrix model, which corresponds
to $q=2$, $\omega(z)$ satisfies a second degree equation. More
precisely, setting $p=2m+1$, one finds
$$\omega(z)={m+1\over\pi}\int^{u(x)}_{-z} {\d u'\, u'{}^m\over\sqrt{z+u'}},$$
and thus
$$\omega^2(z)=(z+u)P^2_{m}(z,u),$$
where $P_{m}$ is proportional to the polynomial part of the large $z$
expansion of $z^m(1+u/z)^{-1/2}$ \refs{\rGZlob,\rGZaplob}.\par
Analogous properties for $q>2$ are maybe less well-known.
\medskip
{\it Operators of third order: $q=3$.}
A short calculation shows that the trace $\omega(z)$ of the resolvent of the
operator $Q=i(\d^3-(3/4)\{\d,u\})$ is, in the spherical and scaling limit,
proportional, up to a rescaling of $x$, to the expression \eomscg\ found in
section2 2.4,2.5. The result relies, in particular, on the property that $p$
which is even and relatively prime with $q=3$, is necessarily of the form
$p=6m+2$ or $p=6m+4$.
\medskip
{\it Generic situation.} In the general case $q\ge 4$ the explicit functional
form of the operator $Q$ depends on the $(p,q)$ models. In the
semiclassical limit the coefficients $w_i$ in equation \eompq\ become
proportional to $u^i$ with in general $p$-dependent coefficients.
However, it is easy to verify, using for example the string actions \rGGPZ\
in the semiclassical limit, that when $p=(2m+1)q\pm1$, the semiclassical form
of the operator $Q$ is $m$-independent. If for convenience we
change the normalizations
$u(x)\mapsto -2u(x)$, $z\mapsto 2i^q z$, and then set in \eompq:
$$y=2u^{1/2}\cos\varphi,$$
we find for the denominator
$$y^q-q u(x)y^{q-2}+\cdots -2z=2\left(u^{q/2}\cos (q\varphi) -z\right).$$
Taking the residue in $y$ we obtain for $\omega(z)$ the expression
$$\omega(z)\propto \int\d x{u^{1/2}\sin\varphi \over u^{q/2}
\sin(q\varphi)} ,$$
where $\varphi$ is solution of the equation
\eqn\eqqphi{u^{q/2}\cos (q\varphi) =z\,.}
It follows
$$\eqalign{ 2i u^{1/2}\sin\varphi&=\left(z+\sqrt{z^2-u^q}\right)^{1/q}-
\left(z-\sqrt{z^2-u^q}\right)^{1/q},\cr
2i u^{q/2}\sin(q\varphi)&=2\sqrt{z^2-u^q}.\cr}$$
We now use the scaling relation $x=u^{(p+q-1)/2}$ and set $s=\sqrt{z^2-u^q}$.
We find
$$\omega(z)\propto \int^{\sqrt{z^2-a^2}}\d s
\left(z^2-s^2\right)^{(p-q-1)/2q}
\left[(z+s)^{1/q}-(z-s)^{1/q}\right],$$
with $a=u^{q/2}$. In the two situations $p=(2m+1)q\pm 1$ (thus $a=
x^{1/(2m+2)}$ or $a=x^{q/(2mq+2q-2)}$) the integration
yields an algebraic expression which, up to normalizations, agrees with the
results \eVint,\eomscpqa,\eomscpqb\ obtained
for the  corresponding $(p,q)$ critical points of the $O(n)$ model.
Note finally that these special $(p,q)$ models include the unitary family
$(q\pm 1,q)$.
\newsec{Conclusion}
We have exhibited all critical points of the $O(n)$ model on a random lattice
generated by the Feynman diagram technique, in the case $n=-2\cos(\pi p/q)$.
We have calculated, in the
spherical limit using steepest descent, the resolvent and the singular free
energy in the scaling limit. In particular we have found that if we
parametrize $p$ as $p=(2m+1)q\pm l$, $0<l<q$, $\gamma_{\rm
string}=-2l/(p+q-l)$. For $l>1$ this result is surprising since it differs
from what is found for the $(p,q)$ critical points of
multimatrix models: $\gamma=-2/(p+q-1)$. One possible interpretation is that
the operator which, in multimatrix models is coupled to the cosmological
constant, is not present here\footnote{$^*$}{We thank I.~Kostov for this
remark.}. \par
We have finally characterized the large order behaviour of the topological
expansion confirming the expected $2k!$ behaviour already found in other
matrix models. \par
We have then shown that in the special case $p=(2m+1)q\pm 1$ all the results
we have obtained are identical to those found in the corresponding $(p,q)$
string models by orthogonal polynomial techniques, and otherwise they are
different. \par
Moreover the techniques developed here allow to obtain a number
of additional results which will be presented in a separate article
\ref\rEZJ{B. Eynard and J. Zinn-Justin, in preparation.}. In particular we
shall study the effect of other relevant operators, a question which has some
subtle aspects since, as the example of the Ising model reveals, negative
powers of the matrix $S$ have sometimes to be considered. We shall
characterize more precisely the
large order behaviour. Finally we  shall exhibit solutions for generic values
of $n$ between $-2$ and $+2$, and discuss the $n\to \pm 2$ limit.
\bigskip
{\bf Acknowledgement.} While this work was in progress, Kostov communicated
us a preprint \ref\rKSt{I.K. Kostov and M. Staudacher, ``Multicritical
phases of the $O(n)$ model coupled to 2D gravity", SPhT/92-025, RU-92-6.}
where  the same model was discussed, though from a slightly different point
of view.
\listrefs
\bye